\documentclass[adp,a4paper,fleqn%
]{w-art}
\usepackage{times,cite,w-thm}
\theoremstyle{plain}

\theoremstyle{definition}

\usepackage[]{graphicx}
\usepackage{color}
\begin{document}                                                                
\DOIsuffix{theDOIsuffix}                                                        
\Volume{16}                                                                     
\Month{01}                                                                      
\Year{2007}                                                                     
\pagespan{1}{}
\Receiveddate{XXXX}
\Reviseddate{XXXX}
\Accepteddate{XXXX}
\Dateposted{XXXX}



\title[Exact results]{Exact results with the Kotliar-Ruckenstein
  slave-boson representation}


\author[R. Fr\'esard]{Raymond Fr\'esard\inst{1,}%
  \footnote{Corresponding author\quad E-mail:~\textsf{Raymond.Fresard@ensicaen.fr},
            Phone: +33-(0)231 45 26 09,
            Fax: +33-(0)231 95 16 00}}
\address[\inst{1}]{Laboratoire CRISMAT UMR CNRS-ENSICAEN 6508, 6 Boulevard
  Mar\'echal Juin, 14050 Caen Cedex, France}
\author[T. Kopp]{Thilo Kopp\inst{2,}}
\address[\inst{2}]{Center for Electronic Correlations and Magnetism, Institute of Physics, 
                      Universit\"at Augsburg, D-86135 Augsburg, Germany}
    \dedicatory{Dedicated to Ulrich Eckern on the occasion of his 60th birthday}
\begin{abstract}
 Radial slave boson representations have the particular advantage that the
 expectation values of their respective fields are finite even without the
 formal introduction of spurious Bose condensates for each of the bosonic
 fields. The expectation values of the radial (real) fields are in fact  
 to be interpreted as the density of empty or singly occupied sites. Whereas
 the radial representation of the Barnes slave bosons has been investigated
 before, a setup for the functional integral of radial bosonic fields in the
 more physical Kotliar-Ruckenstein representation has not been accomplished to
 date. We implement a path integral procedure with suitable renormalization
 factors for a strongly correlated two-site model which allows to control the
 formal steps in the intricate evaluation, as the results for the partition
 function and the expectation values are known from exact diagonalization for
 such a minimal single impurity Anderson model. The partition function is
 shown to be a trace over a product of matrices local in time and therefore
 can be calculated analytically. Eventually, we establish the scheme for the
 evaluation of correlation functions and thermodynamic properties.  
\end{abstract}
\maketitle                   
\section{Introduction}

%
%

Application-oriented, fundamental properties of correlated electronic systems
have become numerous in recent years, in particular for transition metal
oxides: high-T$_c$ superconductivity (see, e.g., \cite{Malozemoff05,Bed86}),
colossal magnetoresistance (see, e.g.,\cite{Hel93}), transparent conducting
oxides (see, e.g.,\cite{Kawa97}), high capacitance
heterostructures~\cite{Li11} or large thermopower (see, e.g.,\cite{Oht07}), to 
quote a few. In addition, they also entail fascinating phenomena such as
superconductivity at the interface of two insulators \cite{Reyren06}, peculiar
magnetism in low dimensional systems \cite{Eyert08}, high temperature
ferromagnetism in vanadate superlattices \cite{Lud09}, all of them providing
a strong challenge to investigate these systems from the theory side. Yet, it
is fair to say that current theoretical approaches meet with severe
difficulties when studying the models which describe these systems. Indeed,
the tool, which is best mastered (perturbation theory), badly fails when the
Coulomb interaction, both local and non-local, is sufficiently strong, thereby
calling for alternative approaches. The introduction of generalized coherent
states \cite{Perelomov} in a constrained Fock space allows for the setup of a
well-defined functional integral for strong Coulomb interaction
\cite{Tuengler} but these path integrals are difficult to handle on account of
the non-trivial Berry phases. An appealing scheme, which is rather based on
canonical fermionic and bosonic fields, the slave boson representation of
local states, was introduced by Barnes to tackle the single impurity
Anderson model (SIAM) \cite{Barnes}. Later the slave boson approach  has been
extended to a whole series of models
\cite{Coleman,KR,Seco,LiWH,FW,FK,Kotliar-Georges,FKopp08,PavlenkoK05} (for a
review see \cite{FKW}) . 

Notably, Kroha {\it et al.} \cite{Kroha} developed a scheme which guarantees
local gauge invariance in a conserving approximation and allows for Fermi
liquid as well as non-Fermi liquid behavior for the investigated multi-channel
Anderson impurity problem. In a different approach towards local gauge
invariance, Read and Newns demonstrated that the phase of the slave boson
field may be gauged away, at the price of introducing a time dependent
constraint \cite{RN83}. By doing so the slave boson is reduced to a radial
slave boson field. Such fields do not possess dynamics on their own, but only
through their coupling to the fermions.  The radial representation has not
been intensively investigated, and the corresponding literature mostly focuses
on the Barnes representation of the SIAM \cite{RFTKO1,FKopp07}.  

Besides, there seem to be (infinitely) many slave boson 
representations to a particular model. This has been shown to be of advantage
by Kotliar and Ruckenstein when investigating the Hubbard model
\cite{KR}, as they devised their representation in such a way that the
Gutzwiller approximation is recovered in a paramagnetic saddle-point
approximation. This approach is exact in the large degeneracy limit
\cite{FW,FLO02} and  obeys a variational principle in the 
limit of large spatial dimensions where the Gutzwiller approximation and the
Gutzwiller wave function are identical \cite{MET89}. Extensions  
\cite{LiWH,FW,FK,Kotliar-Georges} have been applied successfully to a number
of problems \cite{FRE91,meanfield}, all of them making use of a specific
representation of the kinetic energy operator. As the suggested operator takes
an involved form, which is only defined through a corresponding series
expansion, it could not yet be proven to comply to the normal ordering
procedure, and therefore deserves further study. This is the aim of this
paper.  The correctness of various representations of the kinetic energy
operator and interaction terms will be judged by the exact evaluation of the
resulting functional integral representation of the partition function. We
then illustrate the calculation of the expectation value of the radial slave
boson fields. Being intrinsically phase stiff these fields do not average to
zero, in contrast to canonical bosonic fields. Therefore their finite
expectation value is by no means related to a Bose condensate.

\section{Gauge symmetry and radial slave boson representation}

We investigate an exactly soluble model, the simplified single impurity Anderson model enriched by a
non-local Coulomb interaction term. The hamiltonian reads: 

\begin{equation}\label{eqh1}
{\mathcal H} =  \sum_{\sigma} \left( \epsilon_{\rm c} c^{\dagger}_{\sigma}
  c^{\phantom{\dagger}}_{\sigma} + \epsilon_{\rm a} a^{\dagger}_{\sigma}
  a^{\phantom{\dagger}}_{\sigma} + V
  \left(c^{\dagger}_{\sigma}a^{\phantom{\dagger}}_{\sigma} + {\rm
  h.c.}\right)\right) +  U 
a^{\dagger}_{\uparrow}  a^{\phantom{\dagger}}_{\uparrow}  
a^{\dagger}_{\downarrow}  a^{\phantom{\dagger}}_{\downarrow}  
+ I n_{\rm a}n_{\rm c}
\end{equation}
which is defined on two sites. The one-body terms involve
the operators $c^{\dagger}_{\sigma}$
($c^{\phantom{\dagger}}_{\sigma}$) and $a^{\dagger}_{\sigma}$
($a^{\phantom{\dagger}}_{\sigma}$). They represent the creation
(annihilation) of  ``band'' and ``impurity'' electrons,
with spin projection $\sigma$. The energies $\epsilon_{\rm c}$ and
$\epsilon_{\rm a}$ are the band and impurity energy levels, respectively,
while the hybridization energy is $V$. Regarding the two-body terms,  $I$ represents the non-local Coulomb interaction and $U$ is the  on-site repulsion. It
is the largest energy scale in the model and it will be set to infinity
as in the standard infinite-$U$ SIAM. Standard perturbation theory in the
interaction may not be considered for this problem.
While this model may be solved by means of exact
diagonalization \cite{FKopp08}, we here
investigate it by means of a Kotliar and Ruckenstein slave boson
representation. There, the local physical electron operator $a_{\sigma}$ for the impurity site is 
rewritten as a product of the auxiliary fermionic fields $f_{\sigma}$ and auxiliary bosonic fields 
$e$, $p_{\sigma}$ and $d$,  representing
the empty, singly occupied and the doubly
occupied site, respectively. The partition function reads for $U\rightarrow\infty$:

\begin{eqnarray}\label{Zcan}
   {\mathcal Z} &=& \int_{-\pi/\beta}^{\pi/\beta} \frac{\beta d\lambda^{(1)} }{2
   \pi}  
\prod_{\sigma} \int_{-\pi/\beta}^{\pi/\beta} \frac{\beta d\lambda^{(2)}_{\sigma}}{2
   \pi}
\int \prod_{\sigma} D[c^{\phantom{\dagger}}_{\sigma},c^{\dagger}_{\sigma}]
   D[f^{\phantom{\dagger}}_{\sigma},f^{\dagger}_{\sigma}] \int D[e,e^{\dagger}]
\int \prod_{\sigma}
D[p^{\phantom{\dagger}}_{\sigma},p^{\dagger}_{\sigma}]\nonumber \\
&\times & e^{-\int_0^{\beta} d \tau {\mathcal{L}}(\tau)}
\end{eqnarray}
with the Lagrangian

\begin{equation}\label{eqLtot}
{\mathcal{L}}(\tau) = {\mathcal{L}}_f(\tau) +
{\mathcal{L}}_b(\tau) + {\mathcal{L}}_V(\tau) +
{\mathcal{L}}_{\mathnormal{I}}(\tau) 
\end{equation}
consisting of

\begin{eqnarray}\label{Lcont}
{\mathcal{L}}_f(\tau) &=& \sum_{\sigma}
f^{\dagger}_{\sigma}(\tau) \left(\partial_{\tau} - \mu +
i\lambda^{(2)}_{\sigma}\right)f^{\phantom{\dagger}}_{\sigma}(\tau) \nonumber\\ 
{\mathcal{L}}_b(\tau)
&=& e^{\dagger}(\tau)
\left(\partial_{\tau} + i\lambda^{(1)}\right) e(\tau) + 
\sum_{\sigma} p^{\dagger}_{\sigma}(\tau) \left(\partial_{\tau} + i\lambda^{(1)} - i
\lambda^{(2)}_{\sigma} \right) p^{\phantom{\dagger}}_{\sigma}(\tau) - i\lambda^{(1)} \nonumber\\ 
{\mathcal{L}}_V(\tau) &=& V \sum_{\sigma} 
\left(e^{\dagger}(\tau) p^{\phantom{\dagger}}_{\sigma}(\tau)
f^{\phantom{\dagger}}_{\sigma}(\tau) c^{\dagger}_{\sigma}(\tau) + 
e(\tau) p^{\dagger}_{\sigma}(\tau) f^{\dagger}_{\sigma}(\tau)
c^{\phantom{\dagger}}_{\sigma}(\tau) \right) \nonumber\\ 
{\mathcal{L}}_{\mathnormal{I}}(\tau) &=& I \left(1-
e^{\dagger}(\tau) e(\tau)\right) \sum_{\sigma} c^{\dagger}_{\sigma}(\tau)
c^{\phantom{\dagger}}_{\sigma}(\tau) \; .
\end{eqnarray}
The first two contributions entail the dynamics of the auxiliary fields, together with the constraints, specific to the Kotliar and Ruckenstein setup. The hybridization and the non-local Coulomb interaction are represented by the last two terms. Here,
the slave bosons $e$ ($p_{\sigma}$) refer to empty (singly occupied) sites,
respectively. They are subject to the constraints 

\begin{eqnarray}\label{eqrf:q2_KR}
e^\dagger e+\sum_\sigma p^{\dagger}_{\sigma}p^{\phantom{\dagger}}_\sigma &=&1 \nonumber\\
f^{\dagger}_{\sigma} f^{\phantom{\dagger}}_\sigma
-p^{\dagger}_{\sigma}p^{\phantom{\dagger}}_\sigma &=& 0  \;\;\;\;\;\;\;\;\;\;\sigma
=\uparrow,\downarrow \;.
\end{eqnarray}
They are enforced by the
Lagrange multipliers $\lambda^{(1)}$, resp. $\lambda^{(2)}_{\sigma}$. The
Lagrangian, Eq.~(\ref{eqLtot}), possesses a $U(1) \times U(1) \times U(1)$ gauge
symmetry group. Indeed, performing
the gauge transformation of the fermionic fields

\begin{equation}
f^{\phantom{\dagger}}_\sigma(\tau) \longrightarrow e^{-i \chi_\sigma(\tau)} f^{\phantom{\dagger}}_\sigma(\tau)
\end{equation}
and expressing the bosonic fields in amplitude and phase variables as

\begin{eqnarray}
e(\tau) & = & \sqrt{R_e(\tau)} e^{i \theta(\tau)}\nonumber\\
p^{\phantom{\dagger}}_\sigma (\tau) & = & \sqrt{R_\sigma(\tau)}
e^{i (\chi_\sigma(\tau) + \theta(\tau))} 
\end{eqnarray}
allows to gauge away the phases of the three slave boson fields provided one
introduces the three time dependent constraints:

\begin{eqnarray}
\alpha (\tau) & \equiv & i \left( \lambda^{(1)} + \partial_{\tau}
\theta(\tau) \right) \nonumber\\
\beta_\sigma (\tau) & \equiv & i \left( \lambda^{(2)}_{\sigma}
  - \partial_{\tau} \chi_{\sigma}(\tau) \right).
\end{eqnarray}
We here introduced the radial slave boson fields in the continuum limit
following, e.~g., Refs.~\cite{RN83}. When dealing with discretized time
steps special care has to be taken. Extending the procedure introduced in Ref.~\cite{RFTKO1} for Barnes slave bosons to the Kotliar and Ruckenstein representation one obtains the partition function as:

\begin{eqnarray}\label{eq:maz}
{\mathcal Z} &=& \lim_{N\rightarrow\infty} \lim_{\eta\rightarrow 0^{+}}
\lim_{\Omega\rightarrow\infty} \left( \prod_{n=1}^{N}
\int_{-\Omega}^{\Omega} \frac{\delta d\alpha_n}{2 \pi}
\int_{-\Omega}^{\Omega} \frac{\delta d\beta_{\uparrow n}}{2 \pi}
\int_{-\Omega}^{\Omega} \frac{\delta d\beta_{\downarrow n}}{2 \pi}
\int_{-\eta}^{\infty} dR_{e, n}
\int_{-\eta}^{\infty} dR_{\uparrow, n}
\int_{-\eta}^{\infty} dR_{\downarrow, n} \right)\nonumber\\
& \times&  e^{-S_b} {\mathcal Z_f}
\end{eqnarray}
where $R_{e, n}$ ($R_{\sigma, n}$) corresponds to the amplitude of the above
complex $e_n$ ($p_{\sigma, n}$) bosonic variable, $\delta
\equiv \beta/N$, and 

\begin{eqnarray}\label{eq:sb}
S_b &=& i \delta \sum_{n=1}^{N} \left[ \alpha_n\left(R_{e, n}+R_{\uparrow, n} 
+R_{\downarrow, n}-1\right) - \sum_{\sigma} \beta_{\sigma n} R_{\sigma, n} \right] 
\nonumber\\
{\mathcal Z_f} &=& \int \prod_{\sigma} D[c^{\phantom{\dagger}}_{\sigma},c^{\dagger}_{\sigma}]
   D[f^{\phantom{\dagger}}_{\sigma},f^{\dagger}_{\sigma}] e^{-S_f}.
\end{eqnarray}
The fermionic contribution to the action is bi-linear in the fermionic fields. It is given by

\begin{eqnarray}\label{eq:sf}
S_f &=& \sum_{n=1}^{N}\sum_{\sigma} \left\{
f^{\dagger}_{n,\sigma} \left[ f_{n,\sigma} - f_{n-1,\sigma} 
\left[ 1-\delta \left(\epsilon_{\rm a} + i\beta_{\sigma n} -\mu \right) \right] \right] \right. \nonumber\\
&&\left. + 
c^{\dagger}_{n,\sigma} \left[ c_{n,\sigma} - c_{n-1,\sigma}
\left[ 1-\delta \left(\epsilon_{\rm c} -\mu +I(1-y_{\sigma n}^2) \right) \right]\right]\right. \nonumber\\
&&\left. + V \delta 
\left( \bar{z}_{n,\sigma} f^{\dagger}_{n,\sigma} c_{n-1,\sigma} + z_{n,\sigma}
c^{\dagger}_{n,\sigma} f_{n-1,\sigma} \right) \right\} \; .
\end{eqnarray}
Hence the fermions not only experience fluctuating magnetic fields, but the hybridization is effectively fluctuating as well. 
As several forms for the operators $y_{\sigma n}$ and $z_{\sigma n}$ may be
considered we postpone their actual definition. While there is little intuition
linked with the evaluation of Eqs.~(\ref{eq:maz})--(\ref{eq:sf}) the partition
function may be cast in the particularly suggestive form 

\begin{equation}\label{eqh2}
{\mathcal Z} = \lim_{\Omega\rightarrow\infty} \lim_{N\rightarrow\infty} \lim_{\eta\rightarrow 0^{+}}
{\mathcal P} {\mathcal Z_f}
\end{equation}
namely as  a projection of the auxiliary fermion partition function. Here we
introduced 

\begin{equation}\label{eqprojp}
{\mathcal P} = \prod_{n=1}^{N} {\mathcal P}_n
\end{equation}
with

\begin{eqnarray}
{\mathcal P}_n &=& \int_{-\Omega}^{\Omega} \frac{\delta d\alpha_n}{2 \pi}
\int_{-\Omega}^{\Omega} \frac{\delta d\beta_{\uparrow n}}{2 \pi}
\int_{-\Omega}^{\Omega} \frac{\delta d\beta_{\downarrow n}}{2 \pi}
\int_{-\eta}^{\infty} dR_{e,n}
\int_{-\eta}^{\infty} dR_{\uparrow, n}
\int_{-\eta}^{\infty} dR_{\downarrow, n} \nonumber\\
& \times & e^{-i\delta\left(\alpha_n(R_{e,n}+R_{\uparrow, n}+R_{\downarrow, n}-1)-\sum_{\sigma}
  \beta_{\sigma n} R_{\sigma, n}\right)}
\end{eqnarray}
which is defined on one time step only. Having written the projection operator in the simple form 
Eq.~(\ref{eqprojp}) helps to evaluate the partition function  Eq.~(\ref{eqh2}). Nevertheless this may not be straightforwardly achieved as the auxiliary fermion partition function consists of all up-spin and down-spin world lines describing fermions in fluctuating fields. Yet, in   Ref.~\cite{FKopp07}, it has been shown that all these world lines may be re-summed  when expressing 
${\mathcal Z_f}$ in the form:

\begin{equation}\label{eqzfk}
{\mathcal Z_f} = 
\left(\mbox{Tr}\prod_{n=1}^N {\mathcal K}_{\uparrow n} \right) \cdot \left(\mbox{Tr}
  \prod_{n=1}^N {\mathcal K}_{\downarrow n} \right) = 
\mbox{Tr} \left( \prod_{n=1}^N {\mathcal K}_{\uparrow n}
\otimes {\mathcal K}_{\downarrow n} \right)
\end{equation}
where the time steps are apparently not mixed. 
The re-writing of the fermionic determinant Eq.~(\ref{eqzfk}) from the action in
Eq.~(\ref{eq:sf}) may be gained by direct calculation and yields a convenient
form to perform the exact projection onto the physical Hilbert space. The
matrices ${\mathcal K}_{\sigma n}$ read: 

\begin{equation}\label{eqK}
{\mathcal K}_{\sigma n}=
\left(\begin{array}{cccc}
1&~&~&~\\
~&\Lambda_{\sigma n}&\delta V \bar{z}_{\sigma n}&~\\
~&\delta Vz_{\sigma n}&L_{\sigma n}&~\\
~&~&~&\Lambda_{\sigma n} L_{\sigma n}
\end{array}\right)\; .
\end{equation} 
We introduced the short-hand notations:

\begin{eqnarray}\label{eq:l-lambda}
\Lambda_{\sigma n} &=& e^{-\delta \left( \epsilon_{\rm c} -\mu
    +I(1-y^2_{\sigma n})\right)} \equiv L_{\rm c} e^{-\delta I(1-y^2_{\sigma n})} \nonumber \\
L_{\sigma n} &=& e^{-\delta \left( \epsilon_{\rm a} -\mu + i \beta_{\sigma n}
  \right)} \equiv L_{\rm a} e^{-i \delta \beta_{\sigma n} } .
\end{eqnarray}
At this stage we are ready to test various forms that may be taken by the
fields $z_{\sigma n}$, $\bar{z}_{\sigma n}$, and $y_{\sigma n}$. Let us
first concentrate on $z_{\sigma n}$ and $\bar{z}_{\sigma n}$. In the standard Barnes representation they are
both given by $\sqrt{R_{e, n}}$, even though, in Cartesian gauge, they should be replaced correctly by
$e^{\dagger}$ (translated by $ e^*_n$) and $e^{\phantom{\dagger}}$ (translated
by $e^{\phantom{*}}_{n-1}$), respectively (see Ref.~\cite{RFTKO1}). In the Kotliar and
Ruckenstein representation the operator form for $z_{\sigma}$ is $z_{\sigma} = 
e^{\dagger} p^{\phantom{\dagger}}_{\sigma} + p^{\dagger}_{-\sigma} d^{\phantom{\dagger}}
$ ($d \rightarrow 0 $ in the limit $U \rightarrow \infty$). In accordance with
the above result for the Barnes representation one would naturally suggest:

\begin{eqnarray}\label{eq:zzbarkr}
z_{\sigma n} &=& \sqrt{R_{\sigma, n-1} R_{e, n}}\nonumber \\
\bar{z}_{\sigma n}&=& \sqrt{R_{e, n} R_{\sigma, n+1}}
\end{eqnarray}
This forms respects the assignment that when an electron hops onto the
impurity its occupancy changes from empty ($ \sqrt{R_{e, n}}$) to singly
occupied ($ \sqrt{R_{\sigma, n+1}}$), while when the electron leaves the
impurity its occupancy changes from singly occupied 
($ \sqrt{R_{\sigma, n-1}}$) to empty ($ \sqrt{R_{e, n}}$). Of course, the
assumption Eq.~(\ref{eq:zzbarkr}) needs to be verified explicitly, for
instance by computing the partition function
Eqs.~(\ref{eqh2}--\ref{eqzfk}). However an exact evaluation turns cumbersome
since several time steps are involved in the matrices 
${\mathcal K}_{\sigma n}$ and a large number of world lines need to be summed
up. Yet, second order perturbation theory in $V$ may be 
carried out, with the result that the correct answer is in fact obtained with the assignments
Eq.~(\ref{eq:zzbarkr}) up to terms
that vanish in the thermodynamic limit. Though encouraging, this result does
not prove the correctness of the representation  Eq.~(\ref{eq:zzbarkr}) and,
at least, higher orders in $V$ should be considered. But, as infinite order
seems to be beyond reach, we here follow another route.

Let us note that, in their original paper, Kotliar and Ruckenstein used a
different form for the $z$-operator. In operator form, for finite $U$, it reads:

\begin{eqnarray}\label{eqzops}
z^{\phantom{\dagger}}_{\sigma} &=&
e^{\dagger} \left(1-p^{\dagger}_{\sigma} p^{\phantom{\dagger}}_{\sigma} - 
d^{\dagger} d^{\phantom{\dagger}} \right)^{-\frac12}
\left(1-e^{\dagger}e^{\phantom{\dagger}} - 
p^{\dagger}_{-\sigma}p^{\phantom{\dagger}}_{-\sigma}\right)^{-\frac12}
p^{\phantom{\dagger}}_{\sigma} \nonumber \\
&+&
p^{\dagger}_{-\sigma}
\left(1-p^{\dagger}_{\sigma} p^{\phantom{\dagger}}_{\sigma} - 
d^{\dagger} d^{\phantom{\dagger}} \right)^{-\frac12}
\left(1-e^{\dagger}e^{\phantom{\dagger}} - 
p^{\dagger}_{-\sigma}p^{\phantom{\dagger}}_{-\sigma}\right)^{-\frac12}
d^{\phantom{\dagger}}
\end{eqnarray} 
In the limit $U \rightarrow \infty$, $d \rightarrow 0 $. Introducing
radial slave boson fields, Eq.~(\ref{eqzops})---jointly with the
corresponding expression for $ z^{{\dagger}}_{\sigma}$---may be transformed
into:

\begin{equation}\label{eq:zwithroots}
\bar{z}_{\sigma n} = z_{\sigma n} = \sqrt{\frac{R_{e,n}}{1-R_{\sigma, n} }}
\end{equation} 

Regarding the hole density that is represented by $y_{\sigma n}^2$, a natural translation is:

\begin{equation}\label{eq:ysimple}
y_{\sigma n} = \sqrt{R_{e,n}} 
\end{equation}
With this form the time steps are neither mixed in the projection
operator Eq.~(\ref{eqprojp}) nor in the fermion partition function
Eq.~(\ref{eqzfk}). Hence one may write the partition function as

\begin{equation}\label{eq:zdek}
{\mathcal Z} = \mbox{Tr} (k^N)
\end{equation} 
with $k= {\mathcal P}_n
\cdot {\mathcal K}_{n\uparrow} \otimes {\mathcal K}_{n\downarrow}$ where $k$ is
$n$-independent, respecting the translational invariance in time of the
model. Therefore we need to project the entries of this matrix. 
Explicitly, the action of this projection on the various contributions
entering Eq.~(\ref{eqzfk}) are found to be: 

\begin{equation}\label{eqproj}
\begin{array}{@{}lll@{}}
{\mathcal P}_n \cdot 1 = 1&{\mathcal P}_n \cdot \Lambda_{\sigma n}
\Lambda_{-\sigma n} = L_{\rm c}^2 &
{\mathcal P}_n \cdot L_{\sigma n} = L_{\rm a} \\
{\mathcal P}_n \cdot z_{\sigma n} = 1 & {\mathcal P}_n \cdot 
\Lambda_{\sigma n} z_{-\sigma n} = L_{\rm c}&
{\mathcal P}_n \cdot L_{\sigma n} z_{-\sigma n} =0\\
{\mathcal P}_n \cdot y_{\sigma n} = 1 & {\mathcal P}_n \cdot 
\Lambda_{\sigma n} L_{-\sigma n} = L_{\rm c} L_{\rm a} e^{-\delta I} &
{\mathcal P}_n \cdot L_{\sigma n} L_{-\sigma n} =0\\
{\mathcal P}_n \cdot \Lambda_{\sigma n} = L_{\rm c} & {\mathcal P}_n \cdot 
\Lambda_{\sigma n} \Lambda_{-\sigma n}L_{-\sigma n} = L_{\rm c}^2 L_{\rm a} e^{-2\delta I}&
{\mathcal P}_n \cdot L_{\sigma n} \Lambda_{\sigma n}  =L_{\rm c} L_{\rm a} e^{-\delta I}
\end{array}
\end{equation}
where $L_{\rm c}$, $L_{\rm a}$, and $L_{\sigma n}$ are defined through the relations in Eq.~(\ref{eq:l-lambda}). We obtain 

\begin{eqnarray}\label{eqk}
k&=& \left(\begin{array}{c}
1 
\end{array}\right) \oplus
\left(\begin{array}{cc}
L_{\rm c}& \delta V\\
\delta V& L_{\rm a}
\end{array}\right) \oplus
\left(\begin{array}{cc}
L_{\rm c}& \delta V\\
\delta V& L_{\rm a}
\end{array}\right) \oplus
\left(\begin{array}{c}
L_{\rm c}L_{\rm a}e^{-\delta I}
\end{array}\right) \oplus
\left(\begin{array}{c}
L_{\rm c}L_{\rm a}e^{-\delta I}
\end{array}\right) \nonumber \\
&\oplus&  \left(\begin{array}{ccc}
L_{\rm c}^2& L_{\rm c}\delta V&L_{\rm c}\delta V\\
L_{\rm c} \delta V& L_{\rm c}L_{\rm a}e^{-\delta I}&0\\
L_{\rm c} \delta V&0&L_{\rm c}L_{\rm a}e^{-\delta I}
\end{array}\right) \oplus
\left(\begin{array}{c}
L_{\rm c}^2L_{\rm a}e^{-2\delta I}
\end{array}\right) \oplus
\left(\begin{array}{c}
L_{\rm c}^2L_{\rm a}e^{-2\delta I}
\end{array}\right) 
\end{eqnarray}
where the direct sum ($\oplus$) relates to the block-diagonal form of $k$ that
follows from labeling its rows according to: $
(1_{\uparrow},1_{\downarrow})\rightarrow 1,\ 
(1_{\uparrow},2_{\downarrow})\rightarrow 2,\ 
(1_{\uparrow},3_{\downarrow})\rightarrow 3,\ 
(1_{\uparrow},4_{\downarrow})\rightarrow 6,\ 
(2_{\uparrow},1_{\downarrow})\rightarrow 4,\ 
(2_{\uparrow},2_{\downarrow})\rightarrow 8,\ 
(2_{\uparrow},3_{\downarrow})\rightarrow 9,\ 
(2_{\uparrow},4_{\downarrow})\rightarrow 11,\ 
(3_{\uparrow},1_{\downarrow})\rightarrow 5,\ 
(3_{\uparrow},2_{\downarrow})\rightarrow 10,\ 
(4_{\uparrow},1_{\downarrow})\rightarrow 7,\ 
(4_{\uparrow},2_{\downarrow})\rightarrow 12$, with $i_{\uparrow}$
($i_{\downarrow}$) labeling the rows of ${\mathcal K}_{n\uparrow}$ 
(${\mathcal K}_{n\downarrow}$) \cite{zero}. Such a $k$ matrix can be
established for a larger cluster, too. 

When expanded to lowest order in
$\delta$, the blocks of $k$ may be 
diagonalized. It is then easily seen that they may be written as
$e^{-\delta(E_\alpha -\mu N_\alpha)}$ with $E_\alpha$ the eigenvalues of the
Hamiltonian Eq.~(\ref{eqh1}). In particular, in the two-particle sector, the eigenvalues 

\begin{equation}\label{eq:eigsing}
E^{(s)}_{\pm} = \frac{1}{2} \left( 3  \epsilon_{\rm c} + \epsilon_{\rm a} + I 
\pm \sqrt{ (\epsilon_{\rm c}  - \epsilon_{\rm a} - I)^2 +8 V^2 } \right) 
\end{equation}
correspond to the singlet states, and the three-fold degenerate eigenvalue

\begin{equation}\label{eq:eigtrip}
E^{(t)} =  \epsilon_{\rm c} + \epsilon_{\rm a} + I
\end{equation}
corresponds to the triplet states.
This exact evaluation of the partition function
allows us to conclude that the functional integral representation of
this SIAM, Eqs.~(\ref{eq:maz}--\ref{eq:sf}), together with 
Eq.~(\ref{eq:ysimple}) is correct. 

Should we use 

\begin{eqnarray}\label{eq:ysqr}
y_{\sigma n} &=& \sqrt{\frac{R_{e,n}}{1+\epsilon - R_{\sigma, n} }} \nonumber \\
z_{\sigma n} &=& \sqrt{\frac{R_{e,n}}{1-R_{\sigma n} }}
\end{eqnarray}
with $\epsilon = 0^+$, instead of Eq.~(\ref{eq:ysimple}), leads us to the same
conclusion.

In contrast, when using

\begin{equation}
y_{\sigma n} = \sqrt{\frac{R_{e,n}}{1 - R_{\sigma n} }}\; ,
\end{equation}
one encounters ill-defined integrals with the evaluation of 
${\mathcal P}_n \cdot L_{\sigma n} \Lambda_{\sigma n}$.
Consequently this form of little use.
Therefore, while various forms for $y$ and $z$ yield the correct $k$-matrix,
Eq.~(\ref{eqk}), they nevertheless need to be carefully verified. They
are equivalent on the operator level, but their respective field representation
involves a discrete time dependence and the time steps  have to be tuned
meticulously so as to render the functional integrals finite. 

Moreover, while there is no preferred form for $y$ and $z$ when
computing the path integral exactly, their precise choice makes a difference
on the level of the saddle-point evaluation \cite{FRE91,meanfield} (for
a recent comparison, see Refs.~\cite{FOKopp08,FKopp08}). These seemingly
formal properties ensure that the approach captures characteristic features of
strongly correlated electrons as the suppression of the quasiparticle weight
and the Mott-Hubbard/Brinkman-Rice transition \cite{BRI70} to an insulating
state at half filling with increasing on-site Coulomb interaction. In fact,
the Kotliar-Ruckenstein approach has been  impressively successful when
compared to numerical simulations: ground state energies~\cite{FRE91} and
charge structure factors show excellent agreement \cite{ZFW}.

\section{Expectation values, correlation functions, and thermodynamics}

Having gauged away the phases of the slave boson fields leaves us with the
amplitudes of the slave boson fields that are gauge-invariant quantities. In
contrast to ordinary Bose fields, the expectation values of which typically vanish
due to phase fluctuations, radial slave boson fields are generically
characterized by finite expectation values that cannot be ``nullified'' by
phase fluctuations. Indeed, they are intrinsically phase stiff. 

In order to illustrate the above let us determine $\langle R_{\sigma, 1}
\rangle $. According to the above it is given by a specific projection of the
fermionic determinant, namely:

\begin{eqnarray}\label{eqrup1}
{\mathcal Z}  \langle R_{\sigma, 1} \rangle  &=& {\mathcal P} R_{\sigma, 1} \mbox{Tr} \left( \prod_{n=1}^N {\mathcal K}_{\uparrow n}
\otimes {\mathcal K}_{\downarrow n} \right) \nonumber \\
&\equiv& \mbox{Tr} \left(k_{\sigma} k^{N-1}\right)
\end{eqnarray}
with

\begin{eqnarray}\label{eqkup}
k_{\uparrow}&=& \left(\begin{array}{c}
0 
\end{array}\right) \oplus
\left(\begin{array}{cc}
0& 0\\
0& 0
\end{array}\right) \oplus
\left(\begin{array}{cc}
0& 0\\
0& L_{\rm a}
\end{array}\right) \oplus
\left(\begin{array}{c}
0
\end{array}\right) \oplus
\left(\begin{array}{c}
L_{\rm c}L_{\rm a}e^{-\delta I}
\end{array}\right) \nonumber \\
&\oplus&  \left(\begin{array}{ccc}
0& 0&0\\
0& 0&0\\
0&0&L_{\rm c}L_{\rm a}e^{-\delta I}
\end{array}\right) \oplus
\left(\begin{array}{c}
0
\end{array}\right) \oplus
\left(\begin{array}{c}
L_{\rm c}^2L_{\rm a}e^{-2\delta I}
\end{array}\right) 
\end{eqnarray}
and

\begin{eqnarray}\label{eqkdown}
k_{\downarrow}&=& \left(\begin{array}{c}
0 
\end{array}\right) \oplus
\left(\begin{array}{cc}
0& 0\\
0& L_{\rm a}
\end{array}\right) \oplus
\left(\begin{array}{cc}
0& 0\\
0& 0
\end{array}\right) \oplus
\left(\begin{array}{c}
L_{\rm c}L_{\rm a}e^{-\delta I}
\end{array}\right) \oplus
\left(\begin{array}{c}
0
\end{array}\right) \nonumber \\
&\oplus&  \left(\begin{array}{ccc}
0& 0&0\\
0& L_{\rm c}L_{\rm a}e^{-\delta I}&0\\
0&0&0
\end{array}\right) \oplus
\left(\begin{array}{c}
L_{\rm c}^2L_{\rm a}e^{-2\delta I}
\end{array}\right) \oplus
\left(\begin{array}{c}
0
\end{array}\right) 
\end{eqnarray}
Carrying out the algebra (at $T=0$) yields:
\begin{eqnarray}\label{eqrup2}
\langle R_{\sigma, 1} \rangle &=& \left(1-\frac{2V^2}{\Delta^2+4 V^2 +
    \Delta\sqrt{\Delta^2+4V^2}} \right) \delta_{{\mathcal N},1} \nonumber \\
&+& \left(1-\frac{4V^2}{(\Delta-I)^2+8 V^2 +
    (\Delta-I)\sqrt{(\Delta-I)^2+8V^2}} \right) \delta_{{\mathcal N},2} + \delta_{{\mathcal N},3}
\end{eqnarray}
where ${\mathcal N}$ is the number of electrons in the system, and $\Delta \equiv
\epsilon_{\rm c} - \epsilon_{\rm a}$. Therefore $\langle R_{\sigma, 1} \rangle $
arises as a specific projection of the fermionic determinant. It    
is non-vanishing for finite values of the particle number, $V$ and $\Delta$,
thereby putting the above conjecture on firm grounds. 

Regarding the correlation functions they may be easily obtained from the above
steps. For instance, the charge autocorrelation function reads:

\begin{equation}\label{eqcf}
{\mathcal Z}  \langle (R_{\uparrow, 1} + R_{\downarrow, 1})(R_{\uparrow, m} +R_{\downarrow, m}) \rangle = \mbox{Tr}
\left( (k_{\uparrow} + k_{\downarrow}) k^{m-2} (k_{\uparrow}+ k_{\downarrow}) k^{N-m} \right) 
\end{equation}
Again, it is obtained as a specific projection of the fermionic determinant.
This holds true for the spin autocorrelation function as well.

\begin{figure}[h!]
\centerline{\includegraphics[width=0.8\hsize]{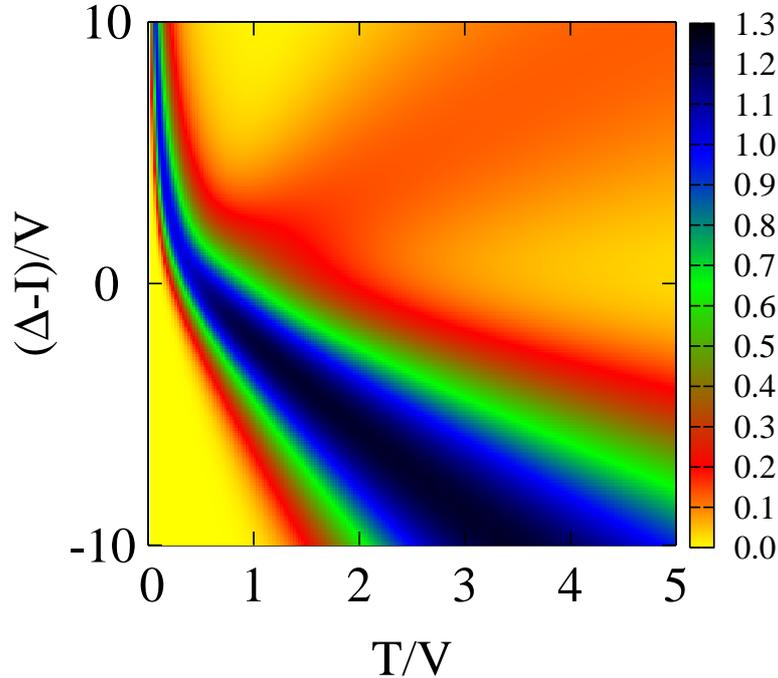}}
\vspace*{-1.5em}
\caption{ (online colour at www.ann-phys.org)
Specific heat $C_{{\mathcal N}=2}$  of the two-site infinite-$U$ SIAM cluster with
nearest neighbor repulsion $I$ and level splitting 
$\Delta =\epsilon_{\rm c}-\epsilon_{\rm a}$. The density plot of $C_{{\mathcal N}=2}$
visualizes the evolution of the specific heat peak with $\Delta-I$.
\vspace*{-1.5em}}
\label{figC}
\end{figure}

With all the energy eigenvalues correctly reproduced,
thermodynamics can be carried through. Remarkably, the specific heat $C_{{\mathcal N}=2}$
only depends on $\Delta -I$. 
For example, for $\Delta -I < 0$, it exhibits a 
Schottky-like shape  for a two-level system, consisting of the levels  $E^{(s)}_{+}$ and $E^{(t)}$ if the single 
particle state with $\epsilon_{\rm a}$ is only thermally occupied and correlations play only a minor role. 
For $\Delta -I > 0$ the specific heat peak is controlled by $(E^{(t)}-E^{(s)}_{-})$, it sharpens significantly, but also 
a weak high-temperature structure at $T \propto (E^{(s)}_{+}-E^{(s)}_{-})$ is observed 
(upper right of  Fig.~\ref{figC}). In the latter regime, correlations play a dominant role 
as the single particle state with $\epsilon_{\rm a}$ is occupied.

\section{Summary}
Summarizing, we have revisited the Kotliar and Ruckenstein slave boson
representation for the $U=\infty$ single impurity Anderson model. It backs on
the introduction of three complex slave boson fields, subject to three
constraints. The $U(1) \times  U(1) \times U(1)$ gauge symmetry
of this representation allows to gauge away the phases of all three
slave boson fields in the continuum limit. We then implemented a path integral
procedure involving radial slave boson fields defined for discrete time
steps. The correctness of several Kotliar and Ruckenstein renormalization
factors has been verified through the exact evaluation of the partition
function and expectation values for a strongly interacting two-site model. In
particular, the expectation value of the radial slave boson fields was shown
to be finite, and does not refer to the introduction of spurious Bose
condensates. On the contrary they directly represent the density of empty 
or singly occupied sites. Correlation functions and
thermodynamics can be correctly captured in this framework. Non-local Coulomb
interaction can be taken into account as well, paving the way for calculations
in the lattice case. For this purpose suitably modified {\it k}-matrices,
which enter the partition function Eq.~(\ref{eq:zdek}), have to be identified. 

{\it Acknowledgments.}\, 
R.F. is grateful for the warm hospitality at the EKM of Augsburg University
where part of this work has been done, and to the R\'egion Basse-Normandie and
the Minist\`ere de la Recherche for financial support. This work was supported
by the Deutsche Forschungsgemeinschaft through TRR 80 (T.K.). Especially, we
would like to thank Ulrich Eckern for his support and encouragement to pursue
our work over so many years. It is a great pleasure to dedicate this paper to
him on the occasion of his 60th birthday.


\begin{thebibliography}{[1]}
\bibitem{zero} In Eq.~(\ref{eqk}) we discarded the remaining $4\times 4$ block
  that vanishes.
\end{thebibliography}
\end{document}